\documentclass{iopart}
\usepackage[english]{babel}
\usepackage{amssymb}
\usepackage[pdftex]{graphicx}

\newcounter{appsect}
\newcounter{speqno}
\newcounter{aeq}[speqno]

\newcounter{aeqno}
\newcounter{sect}
\begin{document}
\newcommand{\p}{\partial}
\newcommand{\B}{\mathbf}
\newcommand{\h}{\hspace{0.1in}}
\newcommand{\f}{\frac}
\newcommand{\ba}{\begin{array}{llll}}
\newcommand{\ea}{\end{array}}
\newcommand{\baq}{\begin{eqnarray}}
\newcommand{\eaq}{\end{eqnarray}}
\newcommand{\be}{\begin{equation}}
\newcommand{\ee}{\end{equation}}
\newcommand{\g}{\gamma}
\newcommand{\dl}{\delta}
\newcommand{\lm}{\lambda}
\newcommand{\s}{\sigma}
\newcommand{\vth}{\vartheta}
\newcommand{\vp}{\varphi}
\newcommand{\G}{\Gamma}
\newcommand{\eps}{\epsilon}
\newcommand{\ra}{\rightarrow}
\newcommand{\Ra}{\Rightarrow}
\newcommand{\Lra}{\Leftrightarrow}
\newcommand{\bel}[1]{\begin{equation}\label{#1}}
\newcommand{\belap}[1]{\begin{appeqa}\label{#1}}
\newcommand{\bsa}[1]{\begin{speqa}\label{#1}}
\newcommand{\esa}{\end{speqa}}
\newcommand{\bsb}[1]{\begin{speqb}\label{#1}}
\newcommand{\esb}{\end{speqb}}
\newcommand{\bsc}[1]{\begin{speqc}\label{#1}}
\newcommand{\esc}{\end{speqc}}
\newcommand{\bsd}[1]{\begin{speqd}\label{#1}}
\newcommand{\esd}{\end{speqd}}
\newcommand{\bse}[1]{\begin{speqe}\label{#1}}
\newcommand{\ese}{\end{speqe}}
\newcommand{\bsap}[1]{\begin{speqap}\label{#1}}
\newcommand{\esap}{\end{speqap}}
\newcommand{\bsbp}[1]{\begin{speqbp}\label{#1}}
\newcommand{\esbp}{\end{speqbp}}
\newcommand{\bscp}[1]{\begin{speqcp}\label{#1}}
\newcommand{\escp}{\end{speqcp}}
\newcommand{\bsdp}[1]{\begin{speqdp}\label{#1}}
\newcommand{\esdp}{\end{speqdp}}
\newcommand{\bsep}[1]{\begin{speqep}\label{#1}}
\newcommand{\esep}{\end{speqep}}
\newcommand{\beap}{\begin{appeqa}}
\newcommand{\eeap}{\end{appeqa}}

\newcommand{\bin}[2]{\left(\ba #1 \\ #2 \ea\right)}
\newcommand{\ff}{\protect\hspace{-0.01in} f \protect\hspace{-0.035in} f}
\newcommand{\sm}{\scriptstyle}

\title[Non-linear dynamics with L\'{e}vy flights]{Weakly non-linear 
dynamics in reaction -- diffusion systems with L\'{e}vy flights}
\author{Y Nec$^1$, A A Nepomnyashchy$^1$ and A A Golovin$^2$}
\address{$^1$ Department of Mathematics, Technion - Israel
 Institute of Technology, Haifa, 32000, Israel}
\ead{flyby@techunix.technion.ac.il}
\address{$^2$ Department of Engineering Sciences and Applied
 Mathematics, Northwestern Univesity, Evanston, Illinois, USA}
\date{}

\begin{abstract}
Reaction--diffusion equations with a fractional Laplacian are reduced 
near a long wave Hopf bifurcation. The obtained amplitude equation is 
shown to be the complex Ginzburg-Landau equation with a fractional 
Laplacian. Some of the properties of the normal complex Ginzburg-Landau 
equation are generalised for the fractional analogue. In particular, an 
analogue of Kuramoto-Sivashinsky equation is derived.
\end{abstract}

\submitto{Physica Scripta}

\section{Introduction}

Random processes characterised by L\'{evy} flights have been discovered 
several decades ago. Since then similar processes have been observed in 
numerous natural phenomena: diffusion and advection in fluids 
\cite{Solomon}, in particular turbulent flows \cite{Hansen} and wave 
turbulence \cite{Balk}, motion of animals \cite{Marell}, balance control 
in humans \cite{Cabrera} and even progress of seismic foci 
\cite{Sotolongo}. At the macroscopic limit L\'{evy} flights are modelled 
by a fractional Laplacian operator. General properties of such processes 
are reviewed in \cite{Viswanathan}. Often non-linear kinetics, such as 
(~but not only~) chemical reactions create an intricate interaction with 
the diffusion process, especially at an instability threshold 
\cite{Albano}-\cite{Zumofen}. Thus an equation combining a fractional 
diffusion operator (~a Laplacian fraction $\g/2$ with $1<\g\leq 2$~) and 
non-linear kinetics is the simplest model to capture the basic effects 
of such an interaction. 

Understanding of pattern emergence and formation in normal 
reaction--diffusion systems near a Hopf bifurcation point was achieved 
by means of amplitude and phase diffusion equations 
\cite{Aranson}-\cite{Tanaka}. 
The current work follows the course of reduction of the fractional 
reaction--diffusion model near such a bifurcation in order to obtain 
and study the fractional analogues of complex Ginzburg-Landau 
(~amplitude~) and Kuramoto-Sivashinsky (~phase diffusion~) equations.

\section{Reduction near bifurcation point}
Consider a two species fractional reaction--diffusion system 
\bel{act_inh} \f{\p \B n}{\p t}=
 \left( \ba d_1 & 0 \\ 0 & d_2 \ea \right)
 \left( \ba {\mathfrak D}_{|x|}^{\g_1} n_1 \vspace{0.1in}\\
 {\mathfrak D}_{|x|}^{\g_2} n_2 \ea \right) + \B f(\B n), \ee
where the Laplacian fractional counterpart is of a generally distinct 
order for each species:
\be {\mathfrak D}_{|x|}^\g \B n(x,t)=-\f{\sec(\pi\g/2)}{2\G(2-\g)}
 \f{\p^2}{\p x^2} \int_{-\infty}^\infty
 {\f{\B n(\zeta,t)d\zeta}{|x-\zeta|^{\g-1}}}, \h 1<\g<2,\ee
and $\B n, \B f$ and $d_j$ are the species concentration vector, 
kinetics function and diffusion coefficients, correpondingly.
Suppose there exists a uniform steady state
$\B n_0$ satisfying $\B f(\B n_0)=\B 0$,
in whose close vicinity $\B f$ varies according to a sensitivity matrix
$(\nabla\B f)_{jk}=\p f_j/\p n_k$, $j,k\in\{1,2\}$. Then Hopf 
bifurcation occurs when its trace vanishes. Let us split the matrix as
\be \nabla\B f =\nabla\B f_0+ \eps^2 
 \left(\ba 0 & 0 \\ 0 & \mu \ea\right), \h \mbox{tr}\,\nabla\B f_0=0,
 \, \eps\ll 1, \, 0<\mu\sim O(1), \ee
rescale
\be \B n (x,t)=\B N (\xi,t_0,t_2,\ldots;\eps), \h
 \xi=\dl x, \, t_j=\eps^j t, \, j=0,2,\ldots \ee
and expand
\bel{asymp_exp} \B N \sim \B n_0 +
 \sum_{j=1}^{\infty} \dl_j \B N_j(\xi,t_0,t_2,\ldots), \h
 \dl_j=\dl_j(\eps). \ee
For the leading order reduction only the first slow temporal and spatial 
scales will be used. Substitution (\ref{asymp_exp}) into (\ref{act_inh}) 
and scrutiny of the resulting system lead to the following choice of 
$\dl$ and $\dl_j$. If $\g_1=\g_2=\g$, the scales are $\dl_j=\eps^j$ and 
$\dl^\g=\eps^2$, where the latter ensues by
\bel{Riesz_scale} {\mathfrak D}_{|x|}^\g y(x)=
 \dl^\g {\mathfrak D}_{|\xi|}^\g y(\xi/\dl). \ee
Then at order $O(\dl_1)$ a system of 
linear homogeneous equations at the bifurcation point is obtained:
\be \f{\p \B N_1}{\p t_0}-\nabla \B f_0 \B N_1=0. \ee
Its solution is
\baq \B N_1=A(\xi,t_2) e^{\lm t_0}\B v_1+
 \mbox{c.c.}, \h \lm=i\omega, \\ \nonumber
 \B v_1= \left( \ba 1 \\ u \ea \right), \h
 u=\f{i\omega-\nabla f_{11}}{\nabla f_{12}}, \h
 \omega^2=\det \nabla \B f_0. \eaq
At subsequent orders the system is not homogeneous. At order 
$O(\dl_3)$ secular non-homogeneous terms coerce a solvability 
condition for $A$, alias fracitonal amplitude or Ginzburg-Landau 
equation:
\be \f{\p A}{\p t_2}=\f{\mu}{2} A +
 \left( \f{d_2+d_1}{2}+i\f{d_2-d_1}{2}\f{\nabla f_{11}}{\omega}\right)
 {\mathfrak D}_{|x|}^\g A- s\, A |A|^2, \ee
where $s$ is constant and depends on $\omega$ and derivatives of $\B f$
up to third order at the bifurcation point.
To obtain the normal form rescale the variables as
\baq t_2 \mapsto \f{2}{\mu}\tau, \h
 \xi \mapsto \left(\f{d_2+d_1}{\mu}\right)^{1/\g} x, \h
  A\mapsto \sqrt{\f{\mu}{2|\Re s|}} A, \nonumber \\ 
 \alpha=\f{d_2-d_1}{d_2+d_1}\f{\nabla f_{11}}{\omega}, \h
 \beta=\f{\Im s}{\Re s} \nonumber \eaq
\bel{NF1} \f{\p A}{\p \tau} = A+(1+\alpha i){\mathfrak D}_{|x|}^\g A-
 \mbox{sign}(\Re s)(1+\beta i) A |A|^2. \ee
If however $\g_1<\g_2$, the expansion scales $\dl_j$ remain the same up
to third order, where the anomalous term first appears: $\dl_j=\eps^j$,
$j\in\{1,2,3\}$. The spatial scale choice is according to the
activator exponent $\dl^{\g_1}=\eps^2$. Then
$1+2\g_2/\g_1>3$ and the anomalous term of the inhibitor will be
neglected at order $\dl_3$. Actually, it will appear only at order
$k+1$, where $k$ is the greatest integer satisfying
$1+2\g_2/\g_1\geq k$. The amplitude equation coincides with 
(\ref{NF1}) upon setting $\g=\g_1$ and $d_2=0$. When $\g_1>\g_2$, the 
same holds with $\g=\g_2$ and $d_1=0$. Thus the fractional analogue of 
the complex Ginzburg-Landau equation is
\bel{CGLE} \f{\p A}{\p\tau}=A+(1+\alpha i)
 {\mathfrak D}_{|x|}^{\g}A-\mbox{sign} (\Re s)(1+\beta i) A |A|^2, \h
 \g=\min\{\g_1,\g_2 \}.\ee
This equation was formerly derived in \cite{Tarasov} in the problem of 
nonlinear oscillators' dynamics with long range interactions.
In the current paper the properties of the super-critical version 
(~$\Re s>0$~) are studied.

\section{Symmetry properties}
The normal complex Ginzburg-Landau equation possesses translational 
space $x\ra x+c_x$, time $\tau\ra\tau+c_t$ and phase $A\ra A\, \exp(ic)$ 
symmetry. Moreover, solutions within the class of modulated waves
\bel{coh_sol} A(x,\tau)=B(r)\, e^{i(q\, x-\varpi \tau)}, \h 
 r=x-v\tau, \h q,\varpi,v \in \mathbb R \ee
are connected by a similarity transformation within the family
$(\alpha-\beta)/(1+\alpha\beta)=\mbox{const}$ \cite{Aranson}. With the 
introduction of anomaly (\ref{CGLE}) loses the Galilean invariance, and 
the symmetry within the class of modulated waves is preserved only for 
$v=0$. A solution with given $(\alpha,\beta)$ connects to a solution 
with $(\alpha',\beta')$ by $B=a B'$, $r=b r'$. Then $q'=b\, q$,
\baq a^2 b^\g=\f{1+\alpha'\beta'}{1+\alpha\beta}
 \f{1+\alpha^2}{1+{\alpha'}^2}, \nonumber \\
 \varpi'=\alpha'-b^\g \,
 \f{1+{\alpha'}^2}{1+\alpha^2}(\alpha-\varpi) \nonumber \\
 b^{-\g}=\f{1+\alpha \alpha'+
 (\alpha-\alpha')\varpi}{1+\alpha^2}.\eaq

\section{Variational formulation}
For the special case $\alpha=\beta$ the normal complex Ginzburg-Landau
equation can be obtained by variation of a functional \cite{Aranson}. 
Rotation $ A \mapsto A \exp(-i\beta \tau)$ gives
\bel{eq_prm_CGLE} \f{\p A}{\p \tau}=(1+i\beta)(A+
 {\mathfrak D}_{|x|}^\g A- |A|^2 A). \ee
Then the functional must be adjusted to yield the fractional Laplacian 
operator:
\baq \Upsilon=\int_{-\infty}^{\infty} U(x,\tau)dx, \nonumber \\
 U=-|A|^2+\f{1}{2}|A|^4 -\f{\sec(\pi\g/2)}{2\G(2-\g)}
 \left\{\f{\p A^*}{\p x} \f{\p}{\p x}\int_{-\infty}^\infty
 {\f{A(\zeta)d\zeta}{|x-\zeta|^{\g-1}}}+ \right. \nonumber \\
 \left. \f{1}{2}(1-\g)A \int_{-\infty}^\infty{\f{\p A^*}{\p\zeta}
 \f{\mbox{sign}(x-\zeta)}{|x-\zeta|^\g}d\zeta}+\mbox{c.c.}\right\}+c 
 .\eaq
It is possible to choose the constant $c$ so that $\Upsilon$ converges, 
as the function $A$ is $\g$-fold differentiable. The first variation is
\be \dl \Upsilon=-\int_{-\infty}^{\infty}{(A-A^2 A^*+
 {\mathfrak D}_{|x|}^\g A)\dl A^* dx} +\mbox{c.c.}\ee
Since
\be \f{\p A}{\p \tau}=-(1+i\beta)\f{\dl U}{\dl A^*},  \ee
all solutions of (\ref{eq_prm_CGLE}) decay asymptotically in time:
\be \f{\p \Upsilon}{\p \tau}=
 \int_{-\infty}^{\infty}\f{\p U}{\p \tau} dx=
 -\f{2}{1+\beta^2}\int_{-\infty}^{\infty} \left| \f{\p A}{\p t}
 \right|^2 dx < 0.\ee

\section{Stability of traveling waves}
Traveling waves $A_q(x,\tau)=\rho \exp(i(q x-\varpi \tau)) $
are a sub-class of (\ref{coh_sol}) and comprise an important solution 
family of (\ref{CGLE}). By
\bel{p1} {\mathfrak D}_{|x|}^\g e^{iqx}=-|q|^\g e^{iqx} \ee
it is straightforward to show that
\bel{A_Riesz} \rho^2=1-|q|^\g, \h \varpi=\beta-(\beta-\alpha)|q|^\g. \ee

\subsection{\normalsize\it Spatially homogeneous oscillations}
Linearising (\ref{CGLE}) about $A_0$, i.e. 
\be A=e^{-i\beta\tau}(1+u+iv), \h u,v\in \mathbb R, \h
 |u|,|v| \ll 1, \ee
splitting into a real system and neglecting non-linear terms of $u$ and
$v$,
\bel{L1} \hspace{-0.5in} \f{\p}{\p\tau}\left(\ba u \\ v \ea\right)=
 -2\left(\ba 1 & 0 \\ \beta & 0 \ea\right)\left(\ba u \\ v \ea\right)+
 \left(\ba 1 & -\alpha \\ \alpha & \h 1 \ea\right)
 {\mathfrak D}_{|x|}^\g \left(\ba u \\ v  \ea\right). \ee
The eigenvalues $\lambda$ of a normal disturbance
\be \left(\ba u \\ v \ea\right)=
 \left(\ba u_1 \\ v_1 \ea\right)e^{\lambda \tau+ikx}  \ee
satisfy
\bel{G2} \lambda^2+2\lambda (1+|k|^\g)+|k|^\g
  \left( (1+\alpha^2)|k|^\g+2(1+\alpha\beta) \right)=0. \ee
Since $\lambda_1+\lambda_2=-2(1+|k|^\g)<0$,
the disturbance is unstable if
\be  \lambda_1 \lambda_2=|k|^\g \left( (1+\alpha^2)|k|^\g+
 2(1+\alpha\beta) \right)<0, \ee
which gives a set of unstable wave numbers when $1+\alpha\beta<0$:
\bel{BFc} 0<|k|<k_m, \h k_m^\g=-2\,\f{1+\alpha\beta}{1+\alpha^2}.\ee
Thus the instability domain in the $(\alpha,\beta)$ plane coincides with 
the normal Benjamin-Feir domain.

\subsection{Arbitrary wave}
The evolution of a small perturbation $a(x,\tau)$ about an arbitrary 
wave solution $A_q$ is governed by
\be  \f{\p a}{\p \tau}=a+(1+i\alpha){\mathfrak D}_{|x|}^\g a -
 (1+i\beta)(2a |A_q|^2+a^* A_q^2).  \ee
Without loss of generality the base wave may be taken one-dimensional. 
However the disturbance should combine a longitudinal and 
transverse waves:
\be a=A_{q+k}(\tau) e^{i(q+k_x)x+ik_y y}+
 A_{q-k}(\tau) e^{i(q-k_x)x-ik_y y}.\ee
The resulting system of equations is
\be \f{d}{d\tau}\left(\ba A_{q+k} \\ A_{q-k}^* \ea\right)=
 {\mathfrak A}_q \left(\ba A_{q+k} \\ A_{q-k}^*  \ea\right), \ee
wherein ${\mathfrak A}_q$ is a $2\times 2$ matrix whose entries depend 
on $k_x$ and $k_y$. With $A_{q\pm k}=A_\pm \exp((\lm \mp i\varpi) \tau)$
the eigenvalues must satisfy a quadratic equation
\bel{Aqk_det} \det\left(\ba \lm - i\varpi -{\mathfrak A}_{11} &
 \h\h -{\mathfrak A}_{12} \\ \\ -{\mathfrak A}_{12}^* &
 \lm + i\varpi -{\mathfrak A}_{22} \ea\right)=0 .\ee
Below some particular relations between the longitudinal and 
tranverse disturbance wave numbers are considered, and (\ref{Aqk_det}) 
is expanded appropriately at the corresponding limits.

Note that the underlying wave $A_q$ is neutrally stable: for $k_x=k_y=0$
the eigenvalues are $\lm_1=0$, $\lm_2=-2(1-|q|^\g)<0$, thus rendering
the long wave disturbances of special interest. For small ratios
$k_x/q, \,\, k_y/q$ one can distinguish between two qualitatively 
distinct cases:
\bel{delta} \ba (i) & O(k_\xi/q) \sim O(k_\eta/q) \sim O(\nu) \\
 (ii) & O(k_\xi/q) \sim O(k_\eta^2/q^2) \sim O(\nu) \ea , \h
 \nu\ll 1. \ee
First suppose that (\ref{delta}$i$) holds. Then the expansion is taken
up to order $O(\nu^2)$ because at $O(\nu)$ the real part of $\lm_1$ 
vanishes:
\bel{lm_1} \hspace{-1in}
 \Re \lm_1 \sim \f{\g}{2} |q|^\g  \left( -(1+\alpha\beta)
 \left( (\g-1)\f{k_\xi^2}{q^2}+\f{k_\eta^2}{q^2}\right) + (1+\beta^2)
 \f{\g|q|^\g}{1-|q|^\g}\f{k_\xi^2}{q^2} \right) + O(\nu^3). \ee
Thus if $1+\alpha\beta<0$, instability is immediate for any wave 
(\ref{A_Riesz}). For $1+\alpha\beta>0$ solving to leading order the 
inequality $\Re\lm_1>0$ yields
\be \left(\f{k_\eta}{k_\xi}\right)^2< \f{1+\beta^2}{1+\alpha\beta}
 \f{\g |q|^\g}{1-|q|^\g} +1-\g, \ee
wherein coercing positiveness of the right-hand side gives a subset 
of unstable wave numbers
\bel{band} q_m<|q| <1 , \h q_m^{-\g}=
 1+\f{\g}{\g-1}\f{1+\beta^2}{1+\alpha\beta}.\ee
Then in the $(k_\xi,k_\eta)$ plane the instability region is bounded by 
two intersecting straight lines.
For traveling waves within this subset pure longitudinal disturbances
(~$k_\eta=0$~) have the highest growth rate, recovering the Eckhaus
instability criterion for the normal Ginzburg-Landau equation. 

Now suppose (\ref{delta}$ii$) holds. In this case an expansion to order
$O(\nu)$ suffices: 
\be \Re \lm_1 \sim -\f{\g}{2}|q|^\g (1+\alpha\beta)\f{k_\eta^2}{q^2}
 +O(\nu^2). \ee
Here the instability ensues only for $1+\alpha\beta<0$.

\section{Phase diffusion equation}
At the opposite limit of small ratios $q/k_x, q/k_y$ no new instability 
criteria emerge. The spatially homogeneous oscillation $A_0$ is unstable 
within the same region $1+\alpha\beta<0$ with respect to disturbances 
(\ref{BFc}). The evolution of perturbations near the domain boundary is 
described by a fractional non-linear phase diffusion equation
(~Kuramoto-Sivashinsky equation fractional analogue~). Define
$0<\eps\ll 1$ so that $1+\alpha\beta=-\eps$. By (\ref{BFc}) the spatial 
coordinate scale is $\chi=\eps^{1/\g} x$.
To find the appropriate temporal scale take $|k|^\g=K \eps$ and expand
(\ref{G2}) in powers of $\eps$. The resulting approximation is
\bel{G2_KS1} \lm_1 \sim \eps^2 \left( K- \f{1}{2}(1+\alpha^2)K^2
 \right) + O(\eps^3). \ee
Hence the temporal scale is $\tau_2 = \eps^2 \tau$.
Using (\ref{Riesz_scale}) and rewriting (\ref{CGLE}) with $\chi$
and $\tau_2$,
\bel{CGLE4KS} \eps^2 \f{\p A}{\p \tau_2}=A+\eps (1-\f{i}{\beta}(1+\eps))
 {\mathfrak D}_{|\chi|}^\g A -(1+i\beta) |A|^2 A. \ee
Taking
\be A=e^{-i\beta\tau_2/\eps^2} r(\chi,\tau_2)
 e^{i\vp(\chi,\tau_2)}, \ee
where
\be r(\chi,\tau_2)=1+\sum_{j=1}^{\infty}\eps^j \,
 r_j(\chi,\tau_2), \h
 \vp=\sum_{j=1}^{\infty}\eps^j \, \vp_j(\chi,\tau_2), \ee
substituting into (\ref{CGLE4KS}), dividing by $\exp(i\vp)$ and using
the expansions
\be e^{\pm i \vp}=1\pm i\sum_{j=1}^{\infty}\eps^j \, \vp_j-\f{1}{2}
 \left(\sum_{j=1}^{\infty}\eps^j \, \vp_j\right)^2 +\cdots\, , \ee
it is possible to collect powers of $\eps$. At order $O(\eps^3)$ the 
following equation is obtained:
\bel{KS_prm} \f{\p \vp_1}{\p \tau_2}=\f{1}{2}(\beta+\f{1}{\beta})
 \left(-\f{1}{\beta}{(\mathfrak D}_{|\chi|}^\g)^2 \vp_1 +
 {\mathfrak D}_{|\chi|}^\g \vp_1^2-2
 \vp_1{\mathfrak D}_{|\chi|}^\g \vp_1 \right) -
 {\mathfrak D}_{|\chi|}^\g \vp_1. \ee
The operator $({\mathfrak D}_{|\chi|}^\g)^2$ is defined in
Fourier space by
\be ({\mathfrak D}_{|\chi|}^\g)^2 e^{iq\chi}=
 |q|^{2\g} e^{iq\chi} \ee
and cannot be related in a simple way to the operator
${\mathfrak D}_{|\chi|}^{2\g}$, because the order $2\g$ exceeds
the range of definition of ${\mathfrak D}_{|\chi|}^\g$.
Note that the coefficients of the linear terms
${\mathfrak D}_{|\chi|}^{\g} \vp_1$ and
$({\mathfrak D}_{|\chi|}^{\g})^2 \vp_1$ are consistent with
(\ref{G2_KS1}$a$).
To bring (\ref{KS_prm}) into parameter independent form define
\be \tau=\tau_2 \, T, \h x=b\, \chi,
 \h \vp=a \, \vp_1. \ee
Then
$$ \f{\p \vp}{\p \tau}=-{\mathfrak D}_{|x|}^\g \vp -
 ({\mathfrak D}_{|x|}^\g)^2 \vp + \f{1}{2}{\mathfrak D}_{|x|}^\g \vp^2-
 \vp{\mathfrak D}_{|x|}^\g \vp, $$
\bel{KS} b^\g=T=\f{2\beta^2}{1+\beta^2}, \h a=\beta+\f{1}{\beta}. \ee
Except for replacing the Laplacian by fractional operators the
resemblance to the Kuramoto-Sivashinsky non-linear phase diffusion
equation is obvious.

\section{Numerical simulations}

The fractional complex Ginzburg-Landau equation (\ref{CGLE}) and 
fractional Kuramoto-Sivashinsky equation (\ref{KS}) have been solved 
numerically by a pseudo-spectral method with time integration in Fourier 
space, Crank-Nicolson scheme for linear operators and Adams-Bashforth 
scheme for nonlinear ones. Periodic boundary conditions, and if 
not specified otherwise, small amplitude random data as an initial 
condition have been used. Due to the extremely diverse dynamics of both 
CGL and KS equations \cite{Aranson,KSbook} the current section is 
limited to the most interesting regimes that show the difference between 
the normal and fractional equations.

In one spatial dimension phase and amplitude turbulence, two regimes of 
complex behaviour known for the normal CGLE \cite{Aranson}, are 
comparable with the anologous dynamics of (\ref{CGLE}). Figure 
\ref{FCGLphaseT} shows phase turbulence regimes for different values of 
$\gamma$ with $\alpha$ and $\beta$ close to the Benjamin-Feir 
instability threshold. As initial condition a spatially homogeneous 
state $Re A=const$, $Im A=const$, $|A|=1$ was used with small amplitude 
random noise added. The upper subfigure corresponds to $\gamma=2.0$, 
i.e. normal CGL equation with its typical picture of phase turbulence 
and small amplitude modulations near $|A|=1$, exhibiting spatio-temporal 
chaos in the form of splitting and merging "cells" \cite{Aranson}. This 
type of dynamics is described by the Kuramoto-Sivashinsky equation (~see 
below~). The middle subfigure shows a phase turbulence regime for 
$\gamma=1.7$. In this case, together with merging and splitting cells,
long-living traveling shocks form, propagating in different directions. 
These shocks "absorb" the adjacent "cells". The lower subfigure 
corresponds to $\gamma=1.5$. Here a single shock forms in the whole 
domain and travels with a constant speed. The shock tails exhibit weak 
chaotic modulations. As shown below, all above regimes can be captured 
by a fractional KS equation, describing phase dynamics near the 
Benjamin-Feir instability threshold. With increasing distance from 
the threshold the shocks self-accelerate and trigger a transition to 
an amplitude turbulence regime, as shown in figure \ref{FCGLphase_amT}.

\begin{figure}[b]
\centering
\includegraphics[scale=0.5]{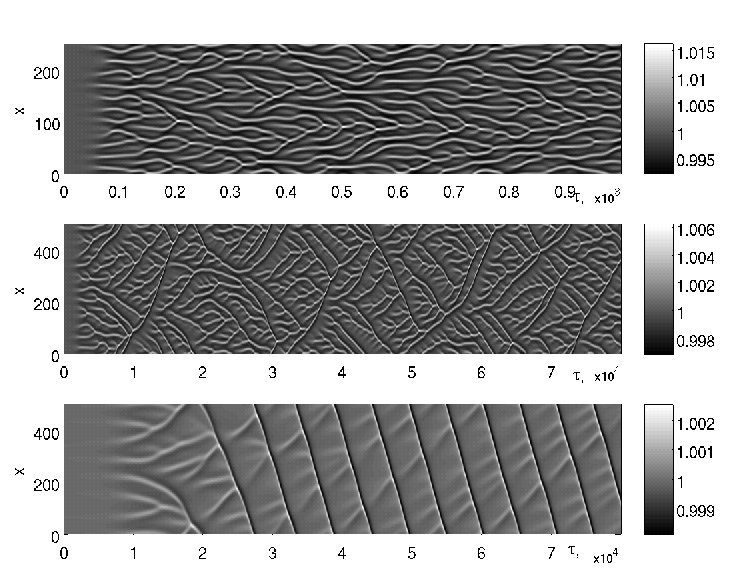}
\caption{Amplitude $|A|$ in a phase turbulence regime as described by 
 FCGL: spatio-temporal diagrams of numerical solutions of (\ref{CGLE}) 
 for $\gamma=2.0,\;\beta=1.2$ (~upper~),
 $\gamma=1.7,\;\beta=1.1$ (~centre~), 
 $\gamma=1.5,\;\beta=1.05$ (~lower~); $\alpha=-1.0$.}
\label{FCGLphaseT}
\end{figure}
\begin{figure}[t]
\centering
\includegraphics[scale=0.5]{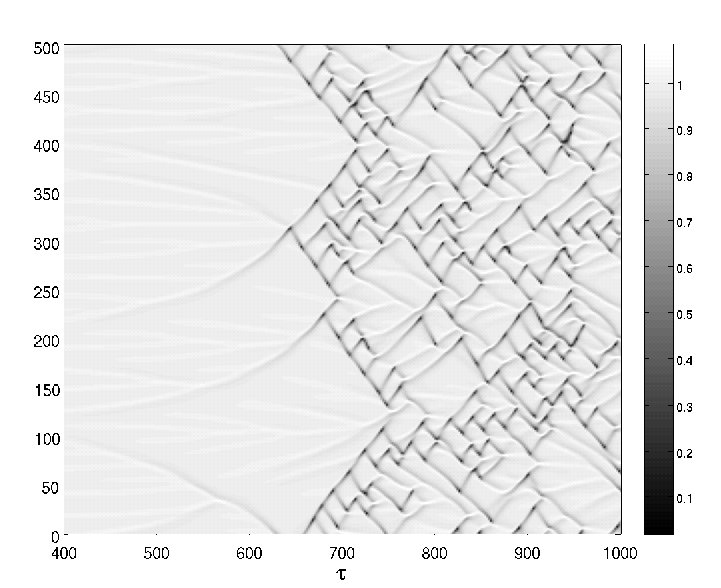}
\caption{Amplitude $|A|$ during transition from phase turbulence to 
 amplitude turbulence triggerred by self-accelerating shocks: 
 spatio-temporal diagrams of numerical solutions of (\ref{CGLE}) for 
 $\gamma=1.6,\;\beta=1.3, \;\alpha=-1.$}
\label{FCGLphase_amT}
\end{figure}

Figure \ref{FCGLamT} shows amplitude turbulence regimes for different 
values of $\gamma$. The usual amplitude turbulence dynamics comprising 
traveling hole solutions with a weak component of the phase turbulence 
is observed for $\gamma=2$ (~upper subfigure; see also \cite{Aranson}~).
As $\gamma$ decreases, the phase turbulence component grows stronger, 
and for $\gamma=1.0$ (~lower subfigure~) a combined phase-amplitude 
turbulent regime is formed with no traveling hole solutions. 

\begin{figure} 
\centering
\includegraphics[scale=0.5]{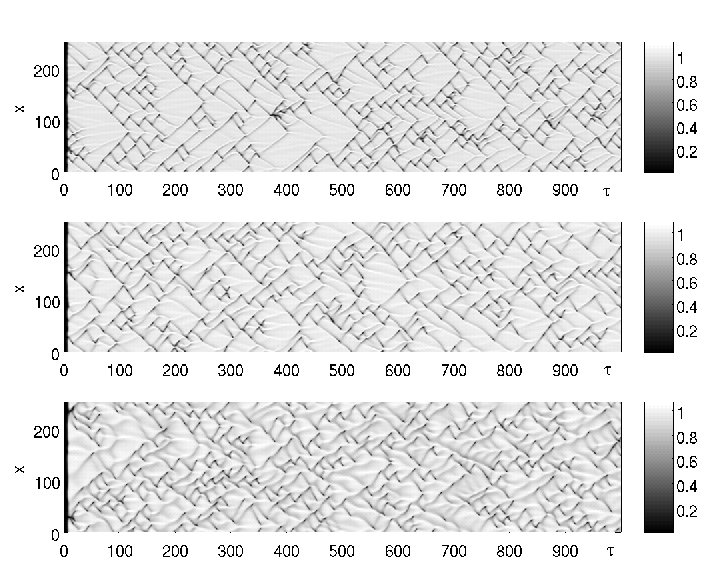} 
\caption{Amplitude $A$ in amplitude turbulence regime as described by 
FCGLE: spatio-temporal diagrams of numerical solutions of (\ref{CGLE}) 
for $\gamma=2.0$ (~upper~), $\gamma=1.5$ (~centre~), and $\gamma=1.0$ 
(~lower~); $\alpha=1.0,\;\beta=-1.3$.} 
\label{FCGLamT}
\end{figure}

Experimental observation of phase and amplitude turbulence regimes 
described by a 1D FCGL equation require special arrangements in order to 
make the experimental system effectively one-dimensional. A more 
generic and easier to study in experiments are two-dimensional systems,
described by a 2D FCGL equation. Probably the most remarkable solution 
of a normal CGL equation in 2D is a spiral wave \cite{Aranson}. 
Therefore the effect of anomalous diffusion on spiral wave dynamics is 
of prime interest.

Numerical simulations of 2D FCGL equation have been performed for 
parameter values, corresponding to the formation of spiral waves in 
normal CGLE, by means of a similar 2D pseudo-spectral code. Figure 
\ref{FCGLspiral} shows snapshots of the solutions for different values 
of $\gamma$. Figures \ref{FCGLspiral}$a,d$ correspond to $\gamma=1.9$. A 
single spiral wave is formed in the whole domain, akin to the normal 
case. In figures \ref{FCGLspiral}$b,e$ (~$\gamma=1.8$~) spiral-like 
defects form, where the core of each defect occupies a certain domain 
with domain walls between them. However, the domain walls in this case 
are partially dissolved and the defects (~spiral cores~) move in a 
chaotic manner. With further decrease of $\gamma$ (~figures 
\ref{FCGLspiral}$c,f$, $\gamma=1.05$~) the number of defects decreases 
and the domain walls between them are almost completely dissolved. The 
system exhibits several domains with almost spatially homogeneous 
oscillations with different phases.

\begin{figure}
\centering
\includegraphics[scale=0.5]{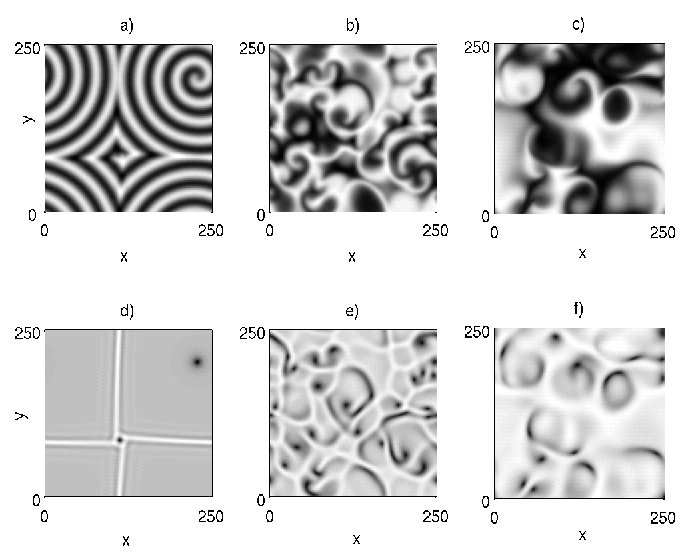}
\caption{Solution snapshots of 2D FCGLE: $\Re A$ (~upper~) and $|A|$ 
(~lower~) for $\alpha=1.5,\;\beta=-0.6$, and $\gamma=1.9$ (a),(d); 
$\gamma=1.8$ (b),(e); $\gamma=1.05$ (c),(f).}
\label{FCGLspiral}
\end{figure}

Finally, the fractional Kuramoto-Sivashinsky equation (\ref{KS}),
describing phase dynamics near the Benjamin-Feir instability threhsold, 
has been solved by a pseudo-spectral code with periodic boundary 
conditions and small amplitude random data as an initial condition. 
Figure \ref{FKS} shows spatio-temporal solution diagrams for different 
values of $\gamma$. Figure \ref{FKS}$a$, corresponding to normal 
diffusion (~$\gamma=2.0$~), shows chaotic spatio-temporal dynamics of 
merging and splitting "cells", as in phase turbulence of the normal CGLE 
shown in figure \ref{FCGLphaseT}(~upper~). Figures \ref{FKS}$b,c$ 
conform to $\gamma=1.7$ and $\gamma=1.6$, respectvely. Along with 
merging and splitting cells traveling shocks appear, absorbing and 
emitting cells. As $\gamma$ decreases, the shocks become more frequent 
and pronounced, and propagate faster, similarly to figure 
\ref{FCGLphaseT} (~centre~). When $\gamma$ decreases below a certain, 
domain length dependent threshold, a single traveling shock is formed in 
the whole domain, as exemplified in figure \ref{FKSshock}: a 
spatio-temporal diagram ($a$) and a few solutions in successive moments 
of time ($b$). The shock travels with a constant speed while its tails 
exhibit weak, chaotic spatio-temporal modulations. A similar shock 
formation is seen in figure \ref{FCGLphaseT} (~lower~).

\begin{figure}[b]
\centering
\includegraphics[scale=0.5]{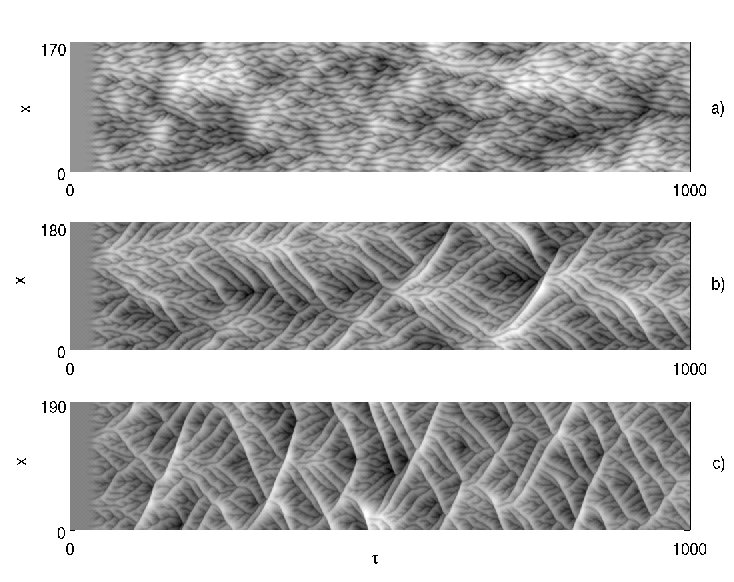}
\caption{Spatio-temporal dynamics of FKS equation (\ref{KS}) for 
 $\gamma=2.0$ (a), $\gamma=1.7$ (b), $\gamma=1.6$ (c).}
\label{FKS}
\end{figure}

Note that the shock amplitude grows both with the decrease of $\gamma$ 
and increase of the computational domain. Below some threshold of $\g$
the shock accelerates with its amplitude growing exponentially. An 
asymptotic analysis revealed large amplitude asymptotic solutions of 
(\ref{KS}) in the form 
\be \vp=a(\tau)f(x-\zeta(\tau)), \ee
where $f$ is an odd periodic function, $a(\tau)$ grows exponentially, 
and the instant speed $d\zeta/d\tau$ is proportional to $a(\tau)$, as
confirmed by numerical simulations. 

\begin{figure}[t]
\centering
\includegraphics[scale=0.5]{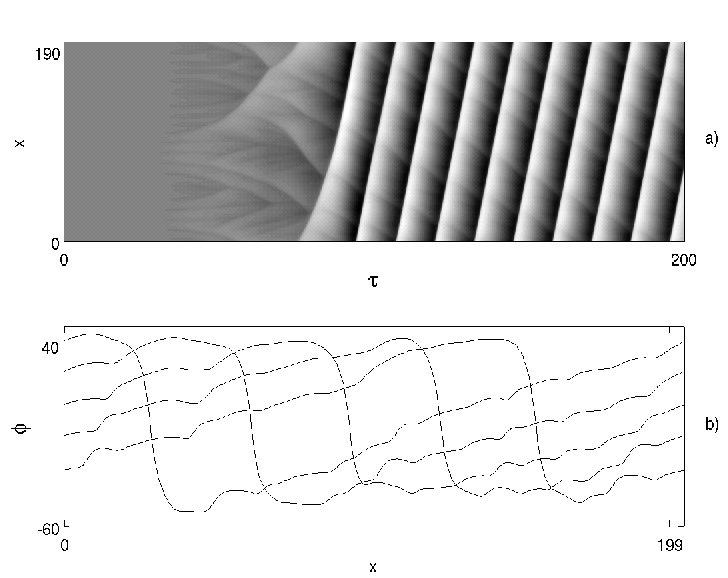}
\caption{Solutions of FKS equation (\ref{KS}) for $\gamma=1.5$:
 spatio-temporal diagram ($a$) and solutions at sucessive moments of 
 time ($b$).}
\label{FKSshock}
\end{figure}

\section{Discussion and conclusions}

A reaction--diffusion system governed by L\'{e}vy flights has been 
reduced near a long wave (~Hopf~) bifurcation point. The reduced system
is described by a complex Ginzburg-Landau (FCGL) equation with a 
fractional order Laplacian.

The fractional analogue does not inherit the Galilean invariance, a 
property well known for a normal Ginzburg-Landau equation, yet the 
similarity of modulated wave solutions along the family of curves 
$(\alpha-\beta)/(1+\alpha\beta)=\mbox{const}$ is retained. Another 
property common to normal and anomalous diffusion is the variational 
formulation in the special case $\alpha=\beta$. 

Similarly to a normal complex Ginzburg-Landau equation, the fractional
analogue possesses the family of solutions in the form of plane 
traveling waves. For a uniform oscillation the instability domain in the 
$\alpha$-$\beta$ space coincides with the normal Benjamin-Feir domain. 
The instability with respect to two-dimensional disturbances ensues for 
the whole unit circle within this domain and for a $\g$-dependent subset 
otherwise.

Near the Benjamin-Feir domain boundary the system dynamics is described 
by a fractional analogue of the Kuramoto-Sivashinsky equation (\ref{KS}) 
for the phase evolution.  

Numerical solution of one-dimensional FCGL equation for decreasing 
values of L\'{e}vy exponent $\g$ reveal the appearance of traveling 
shock waves in the regime of phase turbulence, whereas amplitude 
turbulence exhibit a stronger phase turbulence component. The decrease 
of $\g$ in two-dimensional solutions leads to destruction of spiral 
waves and formation of defect chaos.

Solutions of FKS equation for decreasing $\g$ exhibit a transformation 
of chaotic dynamics of merging and splitting cells, typical of systems 
with normal diffusion, to traveling shocks. The same transition is
observed for the FCGL equation. When $\gamma$ decreases below a certain 
threshold, the shocks self-accelerate, their amplitude grows 
exponentially and the solution blows up. This phenomenon corresponds to 
the transition to amplitude turbulence and has been confirmed by 
numerical simulations.

Some remarks on the possible ways to control the anomalous diffusion 
effects in experiments are in order. For a reaction--diffusion system 
set up in a liquid layer with turbulent mixing the means to control the 
L\'{e}vy exponent, characterising turbulent diffusion, would be to 
control the mixing intensity and possibly the type of turbulent flow. 
For a reaction--diffusion system on a catalytic surface, in which the 
surface super-diffusion would be a consequence of the possibility for 
molecules to make long jumps over the surface through the gas phase, the 
L\'{e}vy exponent might be controlled by temperature change or control 
of the gas flow near the surface by affecting the adsorption bonds, say, 
by light irradiation.

\section*{Acknowledgements}

A.A.N acknowledges the support of Israel Science Foundation grant 
\#812/06. \\ A.A.G acknowledges the support of NSF grant \#DMS-0505878.

\end{document}